\documentclass{epl}
\usepackage{amsmath,epsfig}

\newcommand{\rmd}{{\rm d}}
\newcommand{\bx}{{\mathbf{x}}}

\newcommand{\hMpc}{{\ifmmode{h^{-1}{\rm Mpc}}\else{$h^{-1}$Mpc}\fi}}

\title{Luminosity segregation versus fractal scaling in the galaxy distribution}
\shorttitle{Luminosity segregation versus fractal scaling}
\author{Martin Kerscher\inst{1}}

\institute{
\inst{1} Sektion Physik, Ludwig-Maximilians-Universit{\"a}t, 
Theresienstra{\ss}e 37, D-80333 M{\"u}nchen, Germany
}


\date{draft \today}

\begin{document}
\maketitle

\begin{abstract}
In this letter I present results from a correlation analysis of three
galaxy redshift catalogs: the SSRS2, the CfA2 and the PSCz.  I will
focus on the observation that the amplitude of the two-point
correlation function rises if the depth of the sample is
increased. There are two competing explanations for this observation,
one in terms of a fractal scaling, the other based on luminosity
segregation. I will show that there is strong evidence that the
observed growth is due to a luminosity dependent clustering of the
galaxies.
\end{abstract}

\section{Introduction}

One of the problems in cosmology is to understand the formation of the
large-scale structures in the Universe, as traced by the spatial
distribution of galaxies.  Theoretical models of large-scale
structure and galaxy formation, whether involving analytical
predictions or numerical simulations, are based on some form of random
or stochastic initial conditions.  This means that a statistical
interpretation of the observed galaxy distribution is required, and
that statistical tools must be deployed in order to discriminate
between different cosmological models (e.g.~\cite{martinez:statistics}).
The most frequently employed statistical measure is the two-point
correlation function. Higher-order correlations are important, but
already the observed two-point correlation properties of the galaxy
distribution impose strong constraints on the models of structure
formation.
However, two basically different interpretations of the observed
two-point properties are discussed: in the ``standard'' picture the
galaxy distribution is assumed to be homogeneous on large scales. The
correlations of the deviations from this homogeneous distribution are
quantified by the two-point correlation function $\xi(r)$ (see e.g.\
{}\cite{peebles:lss}).  In the alternative picture the galaxy
distribution is modelled as a fractal. The two-point correlations are
then measured with the conditional density $\Gamma(r)$ (see e.g.\
{}\cite{labini:scale}).  The inhomogeneous nature of a fractal
challenges the standard picture.
Clearly, these two models lead to different interpretations of
observational results.  In this letter I will focus on the growing of the
amplitude of the two-point correlation function $\xi(r)$ with the
sample depth.  This growing amplitude is either explained with the
scaling properties of a fractal or with luminosity segregation, a
luminosity dependent clustering strength.  By reanalysing three galaxy
catalogues I will show that there are strong arguments in favour of
luminosity segregation.

\section{Two-point correlations}

Let me first discuss the stochastic picture where the galaxies
positions in space are treated as a realization of a random process.
The product density $\rho_2(\bx_1,\bx_2)\rmd V(\bx_1)\rmd V(\bx_2)$ is
the probability of finding two galaxies in the volume elements
$\rmd{}V(\bx_1)$ and $\rmd{}V(\bx_2)$, respectively.
In a homogeneous and isotropic point distribution with mean number
density $\rho$ one defines the two-point correlation function
$\xi(r)$ (e.g.~\cite{peebles:lss})
\begin{equation}
\rho_2(\bx_1,\bx_2)= \rho^2(\xi(r)+1) ,
\end{equation}
with $r=|\bx_1-\bx_2|$. The conditional density can
be defined as $\Gamma(r)\equiv\rho(\xi(r)+1)$.
For a point distribution on a fractal the mean number density $\rho$,
and also $\xi(r)$ are not well defined. Thus, in this case, one
investigates the two-point correlations with the conditional density
$\Gamma(r)$: the density of galaxies at a distance of $r$ as seen from
another galaxy {}\cite{pietronero:thefractal}.
A scale invariant cumulant $\xi(r)$ is typically found in critical
systems, whereas a scale invariant conditional density $\Gamma(r)$ is
an indication for a fractal system.  Clearly, only in the limit
$r\rightarrow0$, both $\xi(r)$ and $\Gamma(r)$ may show the same
scaling behaviour.  An instructive discussion of the different scaling
regimes in the galaxy distribution is presented by
{}\cite{gaite:fractal}.

\section{Galaxy samples}

\begin{figure}
\twofigures[width=4.2cm]{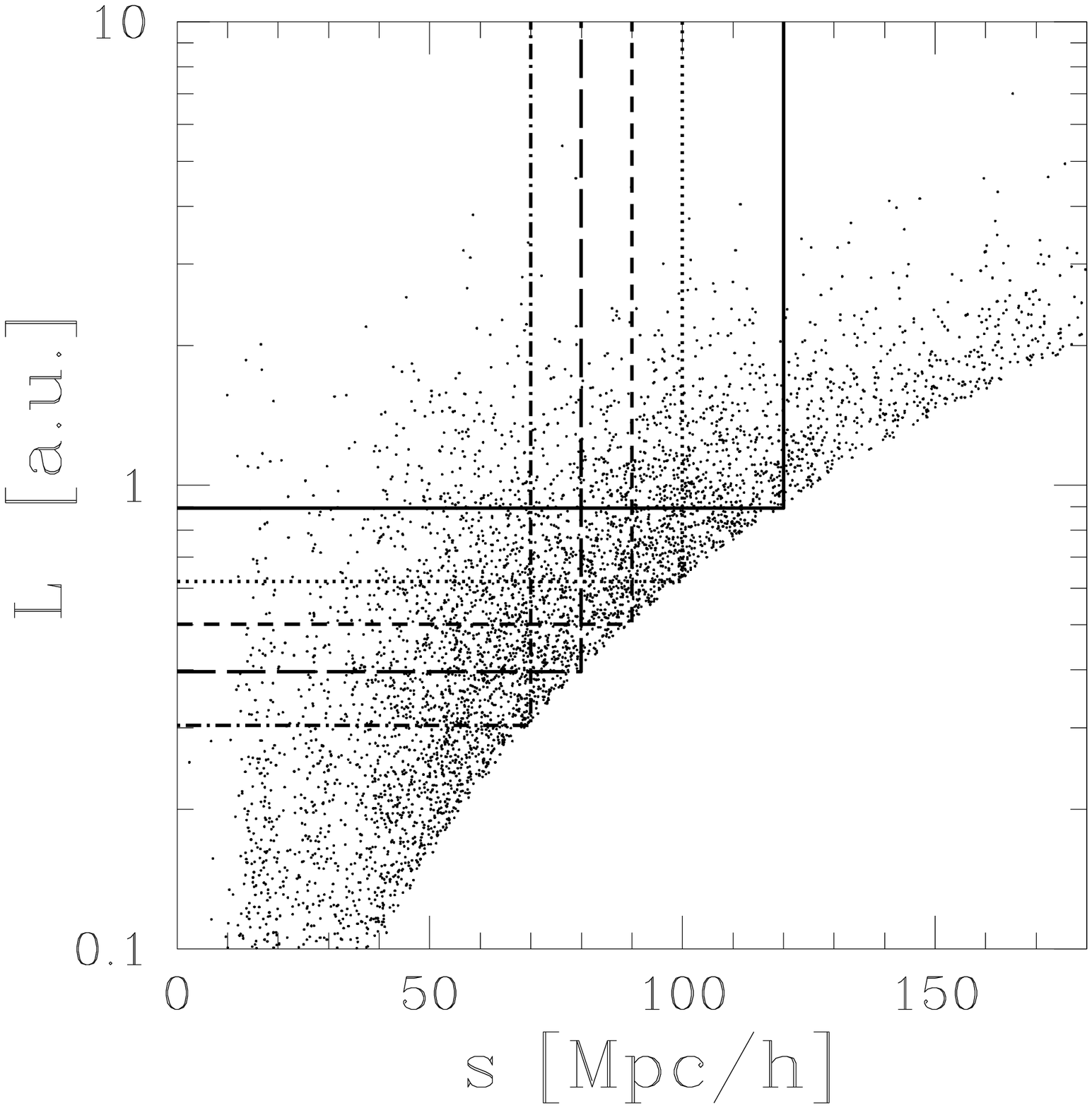}{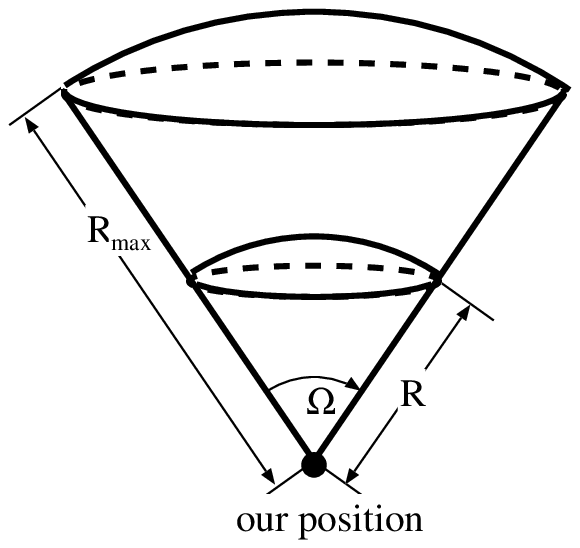}
\caption{Luminosity $L$ of the galaxies in the Southern Sky
Redshift Survey~2 (SSRS2) plotted against their distance $s$ to our
galaxy. Volume-limited samples comprise the galaxies in the upper
left part enclosed by the solid lines (120\hMpc), dotted lines
(100\hMpc), short dashed lines (90\hMpc), long dashed lines (80\hMpc),
and dashed-dotted lines (70\hMpc). Low luminosity galaxies at large
distances are not observed, as can be seen from the empty region in
the lower right part.}
\label{fig:ssrs2-Ls}
\caption{Sample geometry of a volume-limited sample 
with sample depth $R$ (simplified sketch).  Only galaxies inside the opening angle
$\Omega$ and with a distance $s<R$ enter the sample. }
\label{fig:sample}
\end{figure}

In a typical galaxy catalogue the
position on the sky inside a given angular
region $\Omega$ and the flux in a given wave-band are
measured. The distance to our position is estimated from the redshift of
the galaxy. A flux-limited sample consists out of galaxies down to a 
limiting flux $f_{\rm lim}$.
To study the clustering properties of such a galaxy catalogue, I
extract a sequence of volume-limited samples.  A volume-limited
subsample is constructed by introducing a limiting depth $R$ and a
limiting luminosity $L_{\rm lim}$ and by admitting only galaxies
within a distance $s\le R$ from our position and a luminosity $L \ge
L_{\rm lim}$ ($L_{\rm lim}\propto R^2f_{\rm lim}$, in the Euclidean
case, see Fig.~\ref{fig:ssrs2-Ls}).
For a given sample one defines the sample-dependent
number density $\rho_S=N/V$, with the number of galaxies $N$, and the
volume $V$.  The two-point correlation function
$\xi_S$, and the conditional density $\Gamma_S$ determined from this
sample satisfy the relation $\Gamma_S(r) = \rho_S (1+\xi_S(r))$.
If our galaxy sample stems from a homogeneous distribution,
neither $\xi_S$ nor $\Gamma_S$ should change if one increases the
sample size (despite statistical fluctuations).
If our sample stems from a point distribution on a fractal
the number density $\rho_S$ is depending on the sample size, but the
scaling exponent $D-3$ of $\Gamma_S(r)\propto r^{D-3}$ stays invariant 
($D$ is the correlation dimension).  As
a result $\xi_S$ is changing with the size of the sample
(see e.g.~\cite{labini:scale}).  For the two-point correlation
function often the following parameterisation is used
$\xi_S(r)=\left(\frac{r_0}{r}\right)^\gamma$ with the scaling index
$\gamma$. The so-called ``correlation length'' $r_0$ quantifies the
amplitude $r_0^\gamma$ of a scale invariant correlation function.
Consider a large sample with a depth $R_{\text{max}}$ and several
smaller samples $R\le R_{\text{max}}$ within
(see Fig.~\ref{fig:sample}).  On a fractal $r_0$ is proportional to
$R$ {}\cite{coleman:fractal}, specifically
\begin{equation}
\label{eq:fractal_r0}
r_0(R) = \frac{R}{R_{\text{max}}}\ r_0(R_{\text{max}}),
\text{ and }
\xi(r) = 2 \left(\frac{r}{r_0(R)}\right)^{D-3} - 1.
\end{equation}
In the following I will estimate the two-point correlation function
as well as the conditional density for three galaxy samples.  I
checked that for the samples and the scales considered here, 
the estimators discussed by  {}\cite{kerscher:comparison} 
give consistent results.  
I illustrate this by showing the results for $\xi(r)$ both for the
minus (reduced sample) estimator as favoured by {}\cite{labini:scale},
and the estimator due to Landy \& Szalay {}\cite{landy:bias}.
{}\cite{kerscher:comparison} showed that the Landy \& Szalay estimator
has preferable variance properties.

\section{Luminosity segregation but no fractal scaling in the SSRS2}

The Southern Sky Redshift Survey~2 (SSRS2, {}\cite{dacosta:southern})
is 99\% complete with a limiting magnitude of $m_{\rm lim}=15.5$ (magnitudes are 
logarithmic flux measures). The angular extent is $-40^\circ\le\delta\le-2.5^\circ$ with
$b\le-40^\circ$ and $\delta\le0^\circ$ with $b\ge35^\circ$
(declination $\delta$, galactic latitude $b$).
The magnitudes were $K$-corrected as described in
{}\cite{benoist:biasing}, and luminosity distances were used.  Nearly
identical results could be obtained using Euclidean distances and no
$K$-correction.
\begin{figure}
\begin{center}
\epsfig{file=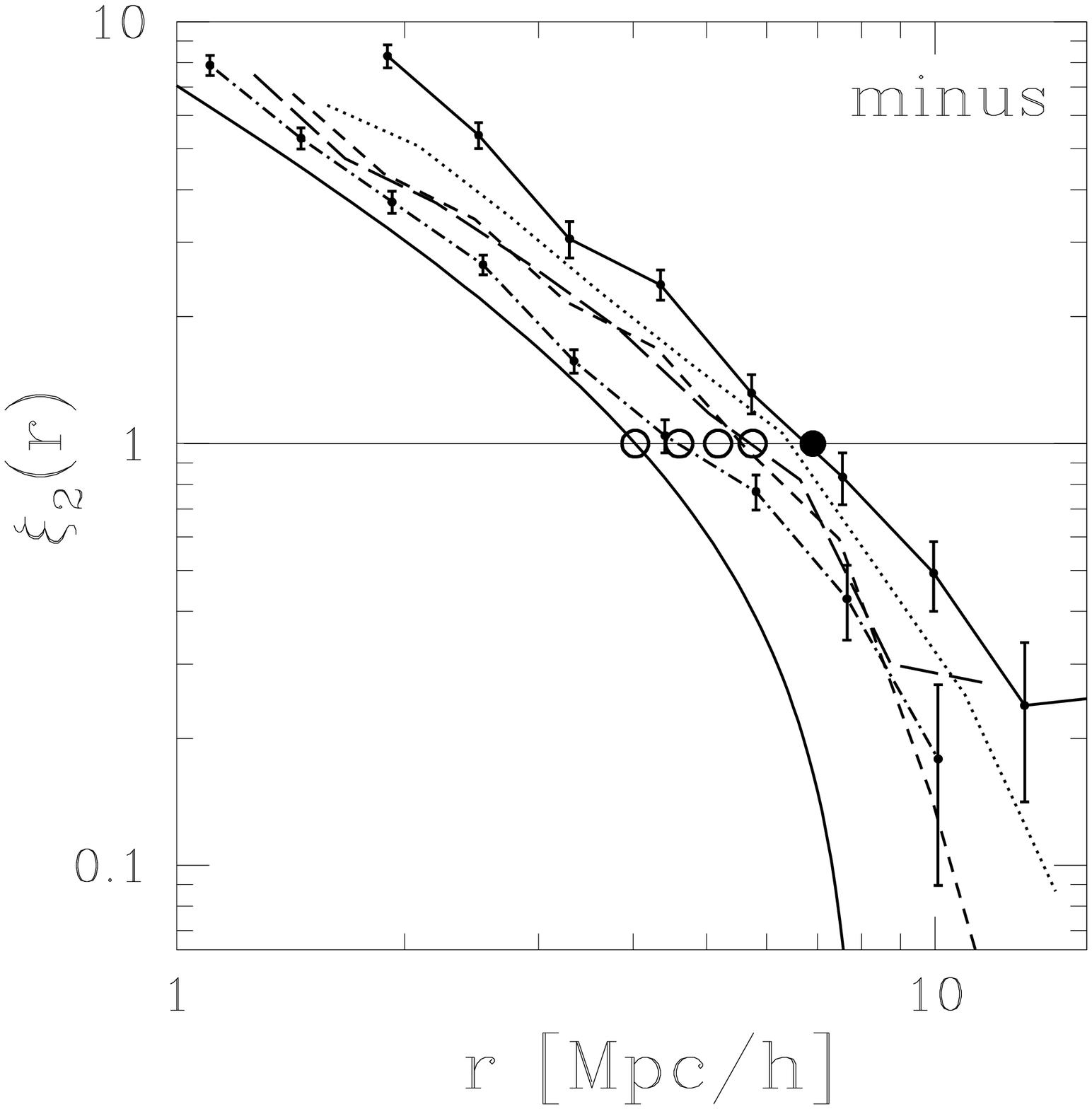,width=4.2cm}
\epsfig{file=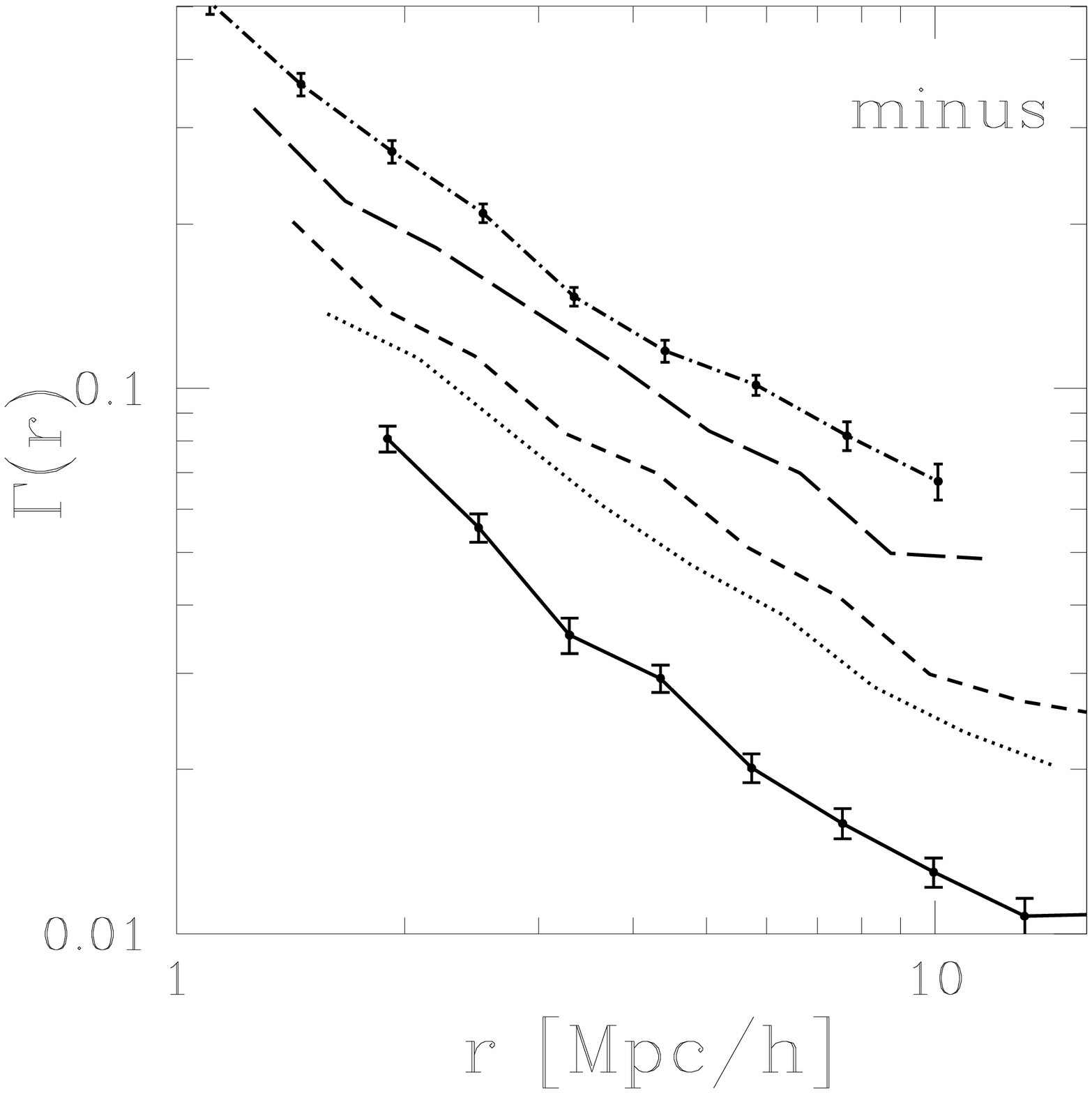,width=4.2cm}
\epsfig{file=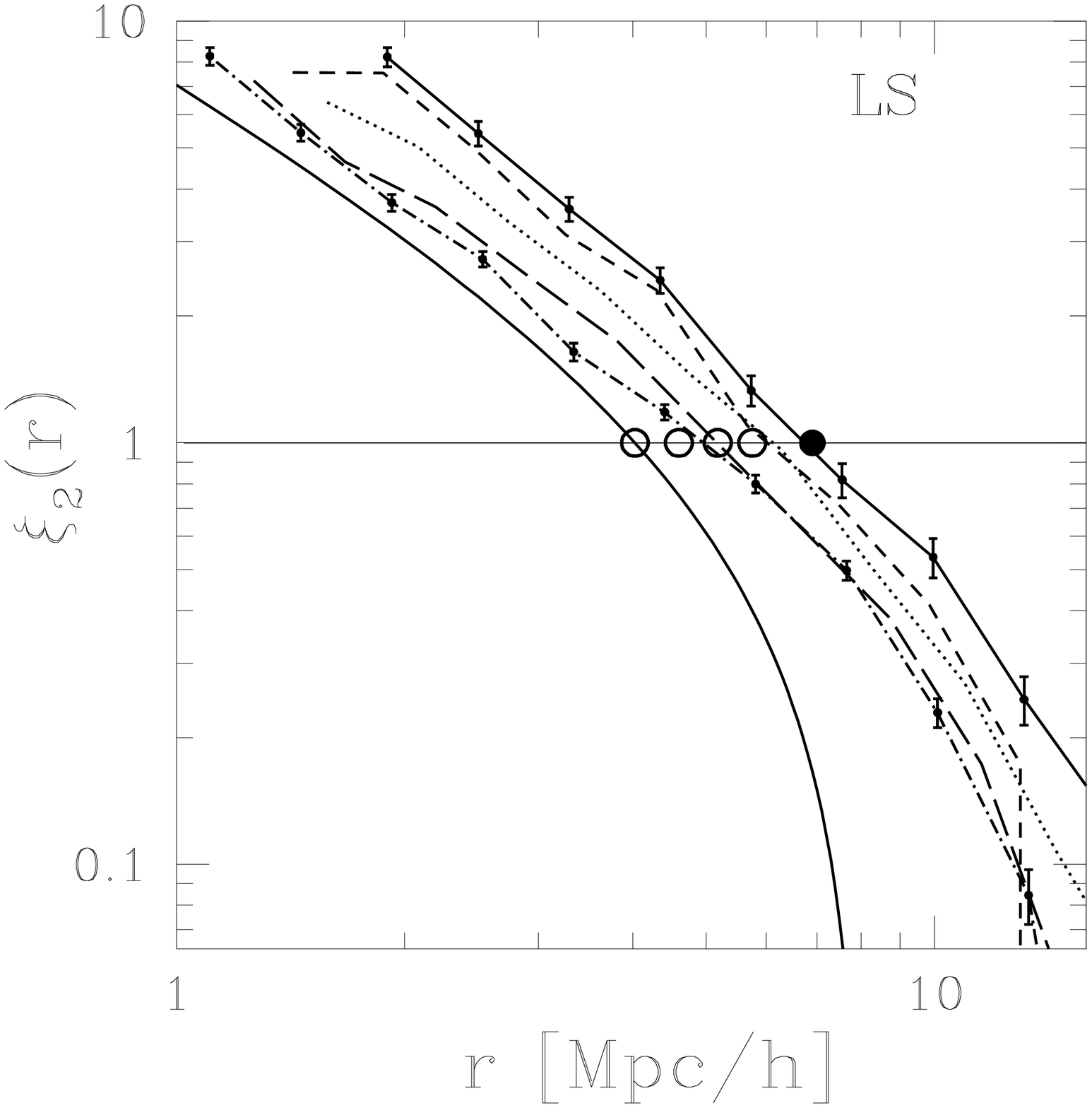,width=4.25cm}
\end{center}
\caption{ $\xi(r)$ and $\Gamma(r)$ for volume limited samples from the
SSRS2 with depth 120\hMpc\ (solid line), 100\hMpc\ (dotted line),
90\hMpc\ (short dashed line), 80\hMpc\ (long dashed line), and
70\hMpc\ (dashed-dotted line). The limiting luminosity changes in these samples.
The labels {\em LS} and {\em minus} mark the results obtained with the Landy \& Szalay and
the  minus estimator, respectively.
The solid dot marks $r_0(120\hMpc)$, the open dots
mark $r_0(100\hMpc)$, $r_0(90\hMpc)$, $r_0(80\hMpc)$, and
$r_0(70\hMpc)$ from right to left, according to
Eq.\eqref{eq:fractal_r0}. The smooth solid line is $\xi(r)$ for a fractal  with $D=2$ 
and $r_0(70\hMpc)$ according to Eq.\eqref{eq:fractal_r0}.
Two-$\sigma$ error bars, determined from a Poisson process, are 
shown only for the 120\hMpc\  and 70\hMpc\ samples.
\label{fig:ssrs2-vollim}}
\end{figure}
In Fig.~\ref{fig:ssrs2-vollim} both $\xi(r)$ and $\Gamma(r)$ are shown
for a sequence of volume-limited samples from the SSRS2. 
The number density in the volume-limited samples decreases from
70\hMpc\ to 120\hMpc.  Consequently, the conditional density
$\Gamma(r)$ is decreasing with the sample depth.
The amplitude of $\xi(r)$ increases with the depth of the samples, and
$r_0$ roughly follows the relation~\eqref{eq:fractal_r0}.  This was
interpreted as a sign of a fractal galaxy distribution (e.g.\
{}\cite{labini:scale}). However, another explanation is possible.  Due
to the construction of the volume-limited sample the mean absolute
luminosity of the galaxies in the sample increases with the depth of the sample
(compare with Fig.~\ref{fig:ssrs2-Ls}).  Hence, the growing amplitude of the
correlation function for deeper volume-limited samples may be a
result of the stronger clustering of the more luminous galaxies. This
is called luminosity segregation.  Clearly, the two-point correlation
function $\xi(r)$ applied to a sequence of volume-limited samples is
not able to distinguish between both claims.
To test the scaling relation~\eqref{eq:fractal_r0} independent from
any luminosity dependence I use a volume-limited sample with a depth
of $R_{\text{max}}=120\hMpc$.  From this volume-limited sample I
extract a sequence of subsamples with depths of $R\le R_{\text{max}}$ (see Fig.~\ref{fig:sample}).
All these subsamples have the same lower limit in luminosity (see
Fig.~\ref{fig:ssrs2-Ls}).
As can be seen from Fig.~\ref{fig:ssrs2-seq} the estimated $\xi(r)$ are
consistent in these samples but inconsistent with the fractal prediction.  
There is no indication for a fractal
scaling of the ``correlation length'' $r_0(R)$ as given in
Eq.\eqref{eq:fractal_r0}.  Moreover, the conditional densities
$\Gamma(r)$ of these samples nearly overlap.
\begin{figure}
\begin{center}
\epsfig{file=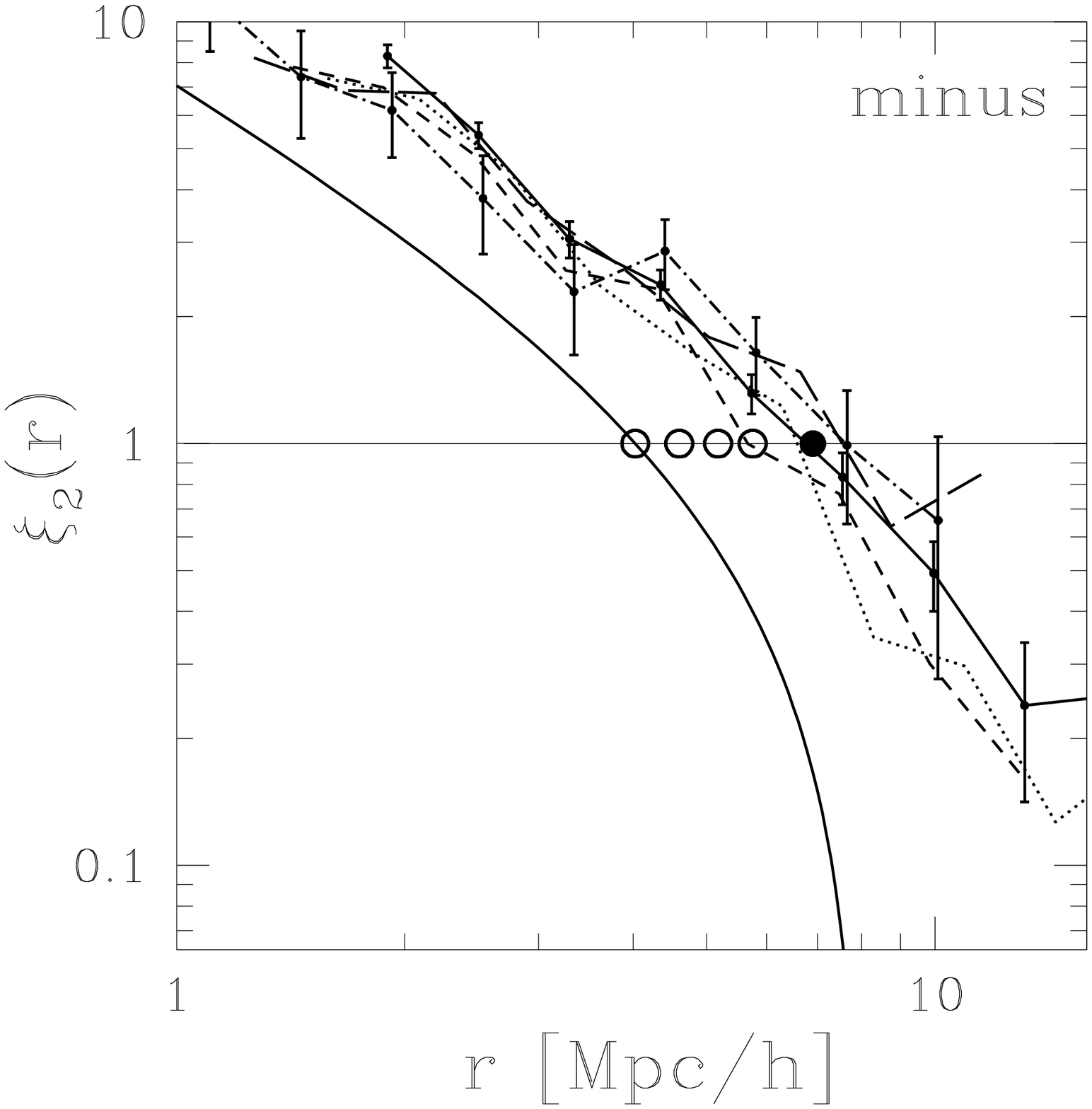,width=4.2cm}
\epsfig{file=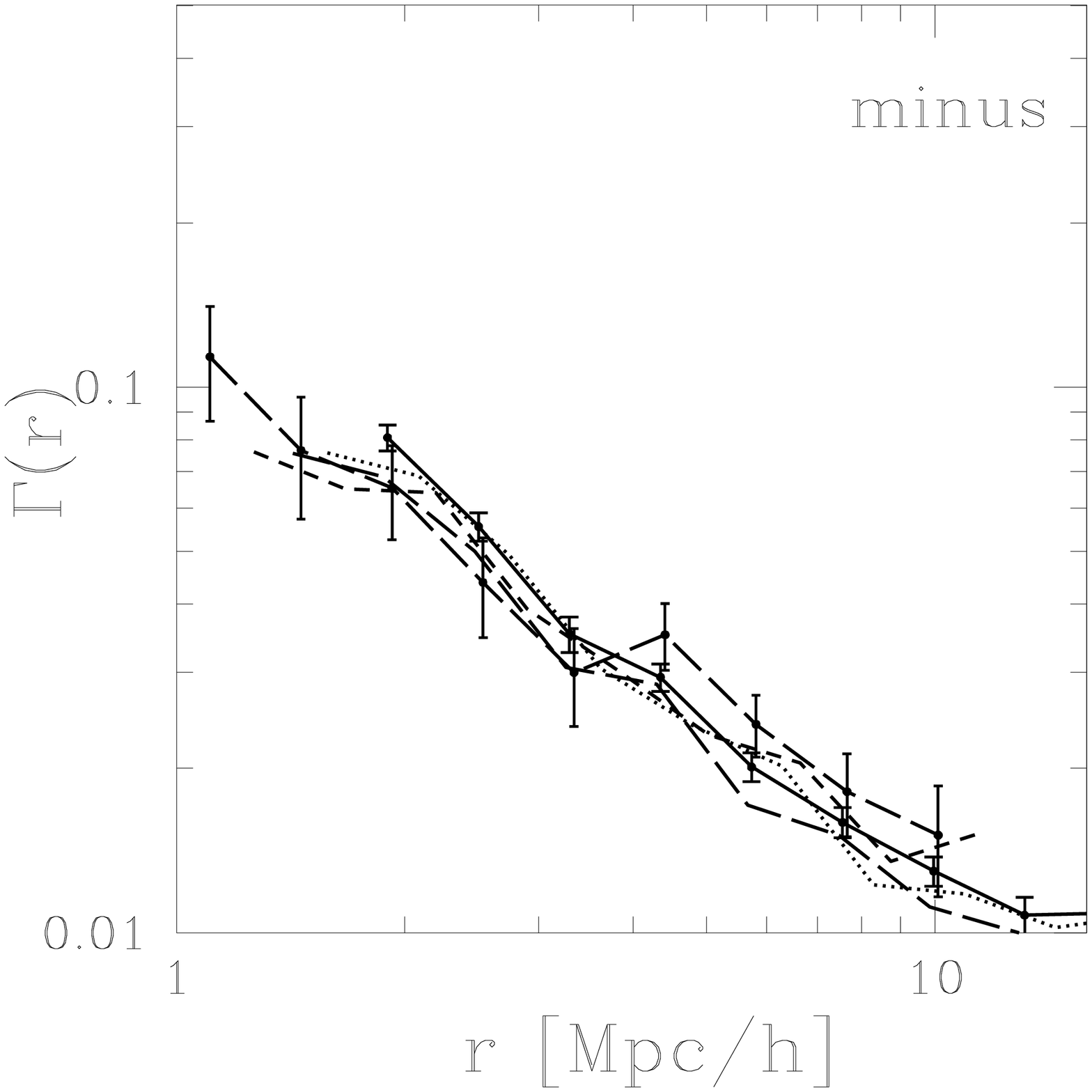,width=4.2cm}
\epsfig{file=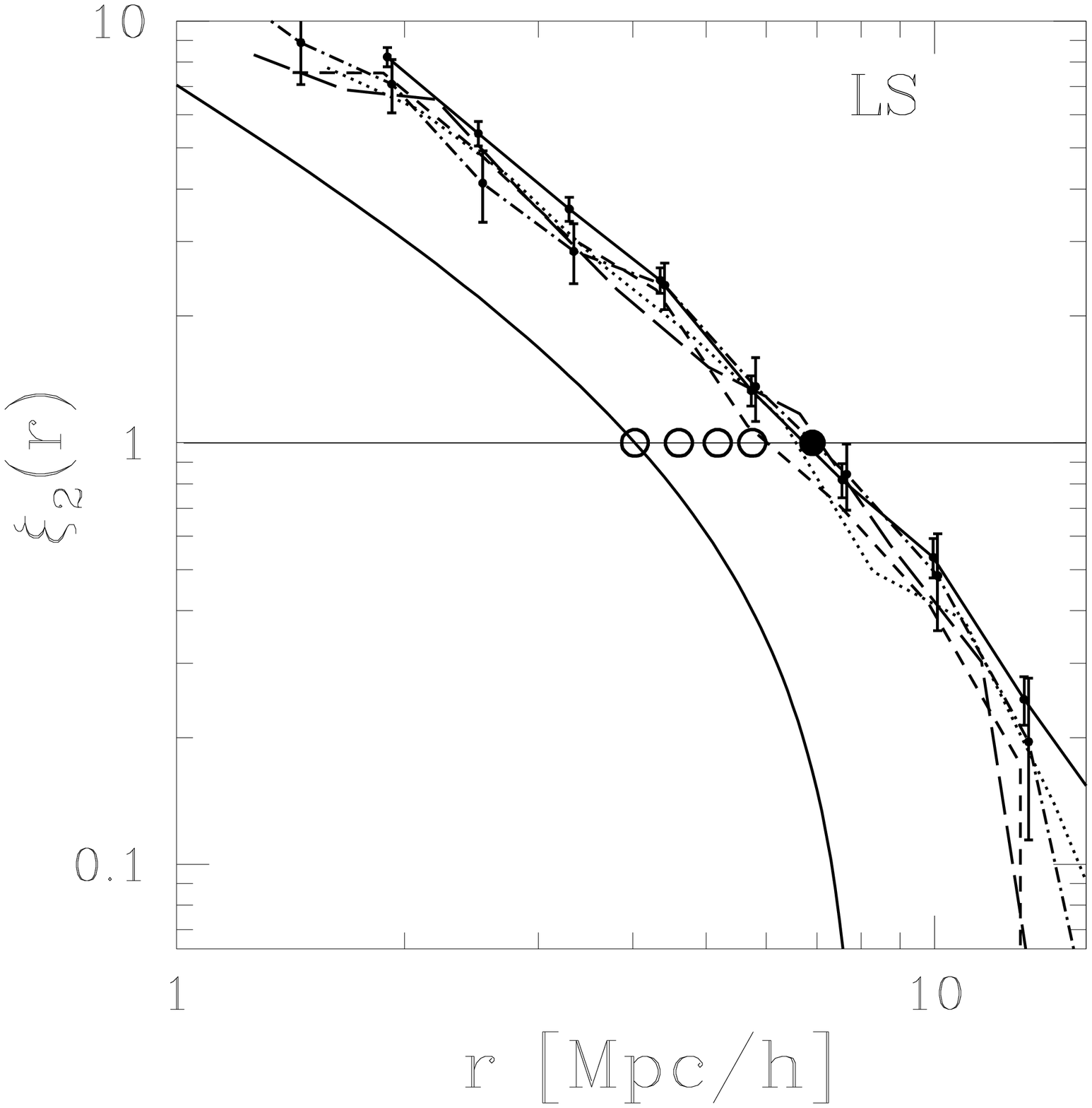,width=4.2cm}
\end{center}
\caption{ $\xi(r)$ and $\Gamma(r)$ for subsamples from the
volume-limited samples from the SSRS2 with a depth of $R_{\text{max}}=120\hMpc$.  The
depth of the sub-samples is 120\hMpc\ (solid line), 100\hMpc\ (dotted
line), 90\hMpc\ (short dashed line), 80\hMpc\ (long dashed line), and
70\hMpc\ (dashed-dotted line). All samples have the same 
limiting luminosity.  The smooth solid line is $\xi(r)$ for a fractal  with $D=2$ 
and $r_0(70\hMpc)$ according to Eq.\eqref{eq:fractal_r0}.
Marks as in Fig.~\ref{fig:ssrs2-vollim}.
\label{fig:ssrs2-seq}}
\end{figure}
Measurement errors, e.g.\ for the position of the galaxies, have no visible 
effect on the correlation functions. The dominant contribution is the statistical error. 
However there is only one realisation of the galaxy distribution in the Universe. 
Therefore, I have to assume a model to quantify the statistical errors.
The simplest model is a purely random distribution of points, the Poisson process. 
I estimate the errors from 100 realisations of a Poisson process with the sample 
geometry, and the number density as in the galaxy samples. 
Later on I will show that a more realistic modelling leads 
to larger errors. These errors are within the same order as determined 
from a Poisson process. For both models, the errors are smaller than the predicted effects 
from fractal scaling.

\section{Luminosity segregation but no fractal scaling in the CfA2}

\begin{figure}
\twofigures[width=4.2cm]{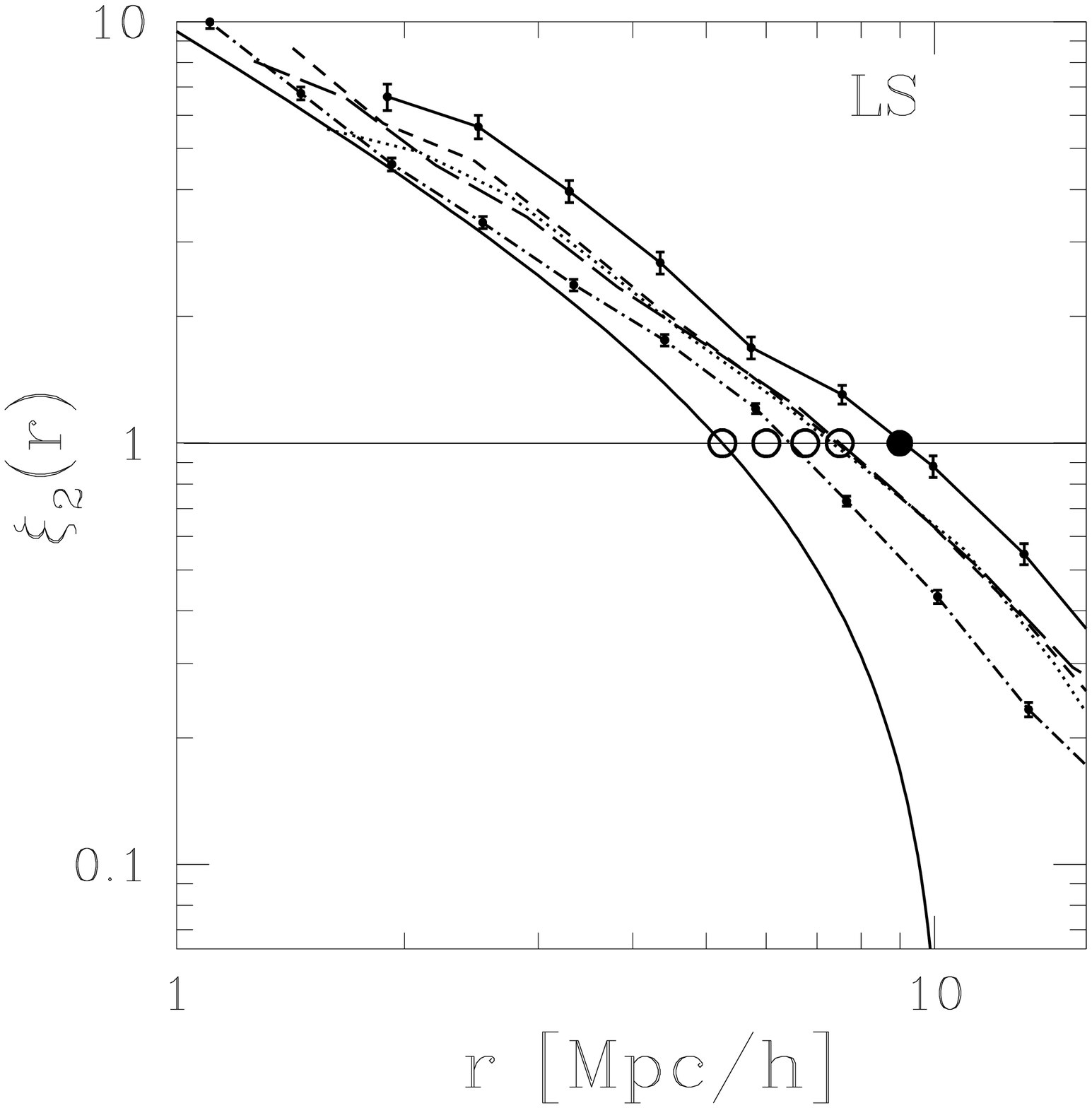}{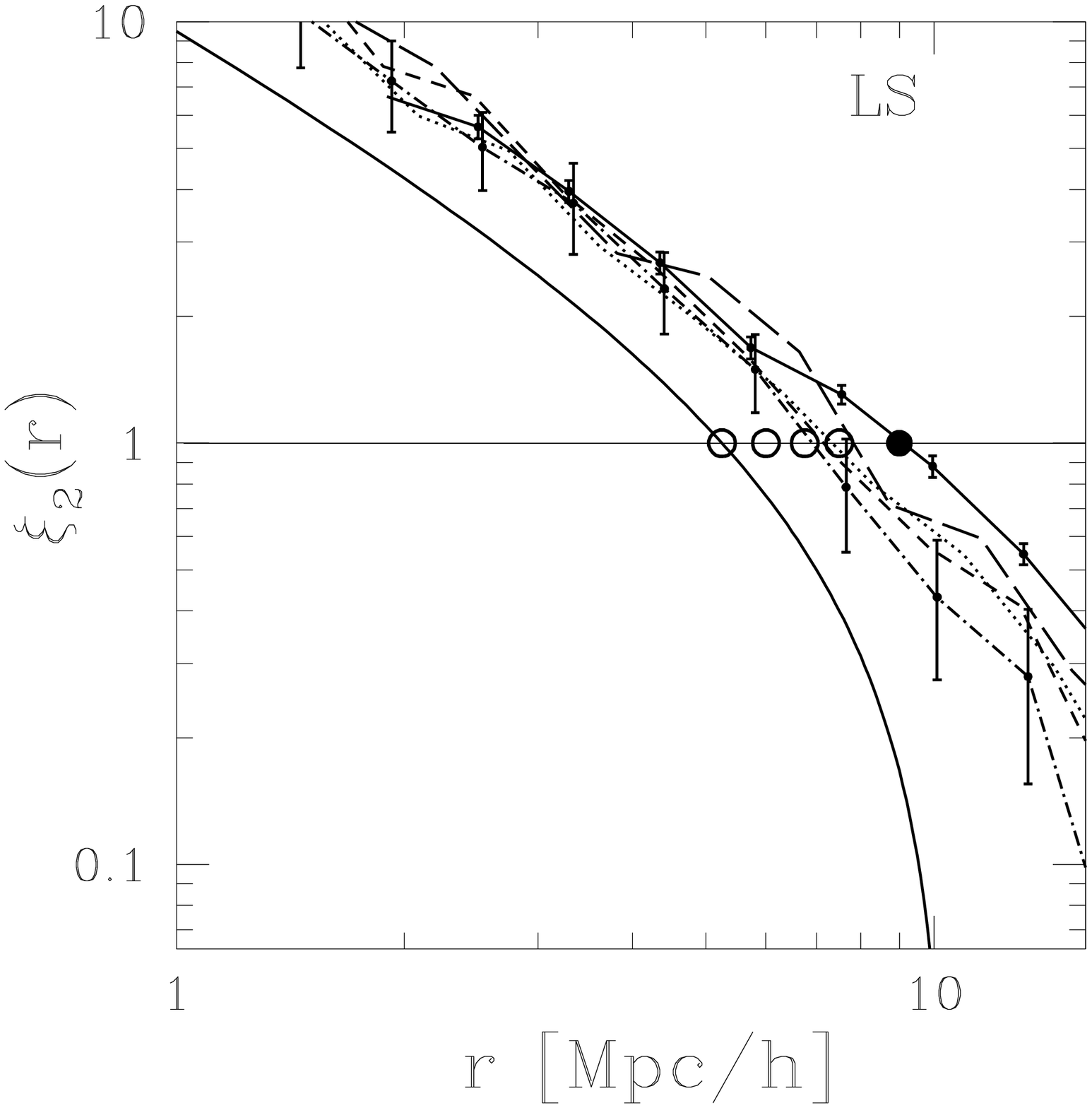}
\caption{$\xi(r)$ for a sequence of volume-limited samples with
changing limiting Luminosity $L_{\rm lim}$ from the CfA2 determined
with the Landy\&Szalay estimator. The smooth solid line is $\xi(r)$ for a fractal  with $D=2$ 
and $r_0(70\hMpc)$ according to Eq.\eqref{eq:fractal_r0}.
Marks as in Figs.~\ref{fig:ssrs2-vollim}.}
\label{fig:cfa2-vollim}
\caption{$\xi(r)$ for samples with varying depth but with the same
limiting luminosity extracted from a volume-limited sample of the
CfA2 with a depth of $R_{\text{max}}=120\hMpc$. Marks as in
Fig.~\ref{fig:ssrs2-seq}. 
\label{fig:cfa2-seq}
}
\end{figure}

As a spatially complementary sample to the SSRS2 I use a galaxy sample
from the CfA galaxy catalogue (see {}\cite{huchra:cfa2south} and
references therein), with $\delta\ge0$, and $|b|\ge35$ and a limiting
magnitude of $m_{\rm lim}=15.5$. From this galaxy catalogue I extract
similar volume-limited samples as for the SSRS2.
The results shown in Figs.~\ref{fig:cfa2-vollim},\ref{fig:cfa2-seq} lead to the
same interpretation as for the SSRS2: the growing amplitude of
$\xi(r)$ is caused by luminosity segregation. 

\section{Neither luminosity segregation nor fractal scaling in the PSCz}
\label{sect:pscz}

\begin{figure}
\twofigures[width=4.2cm]{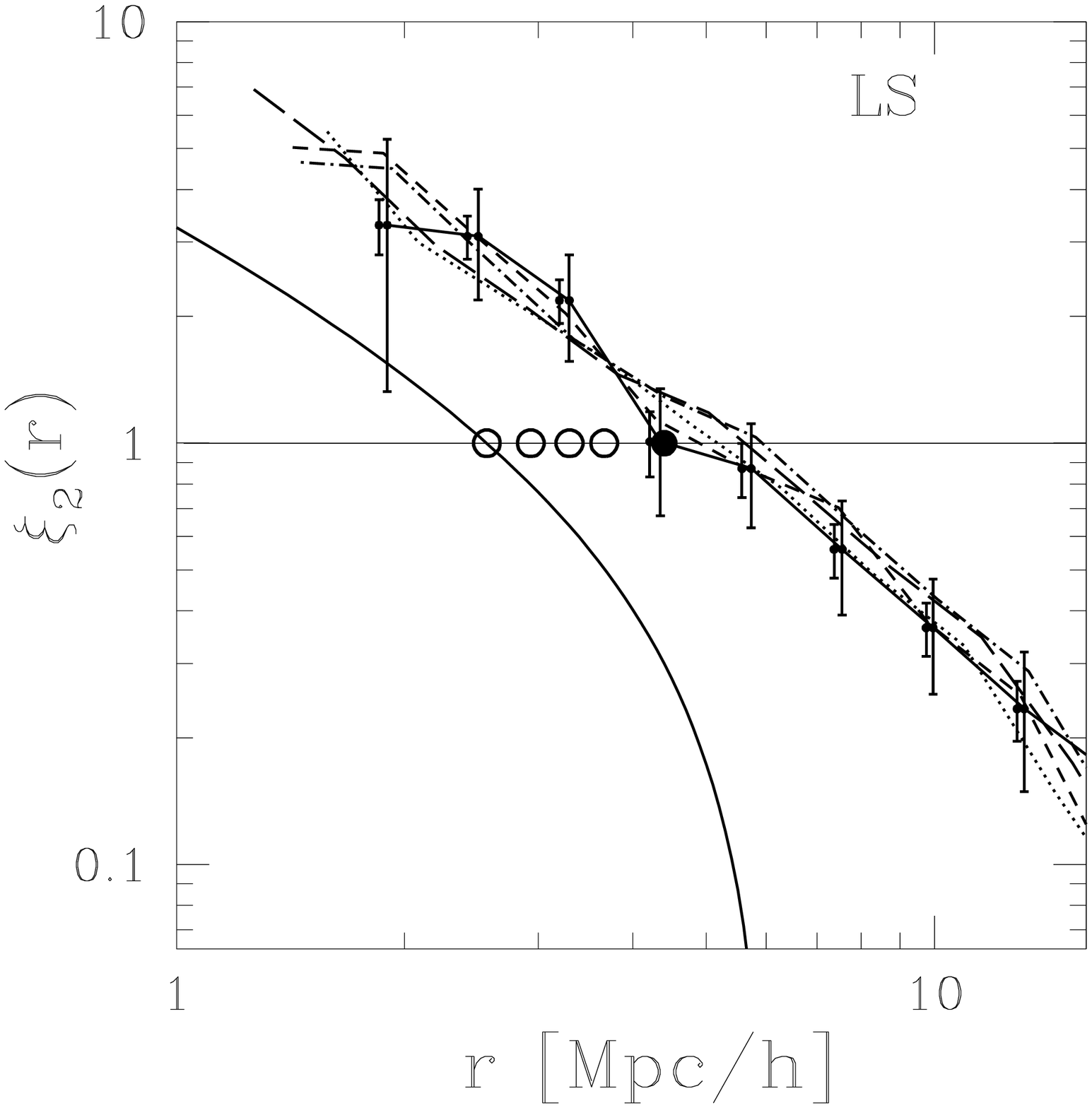}{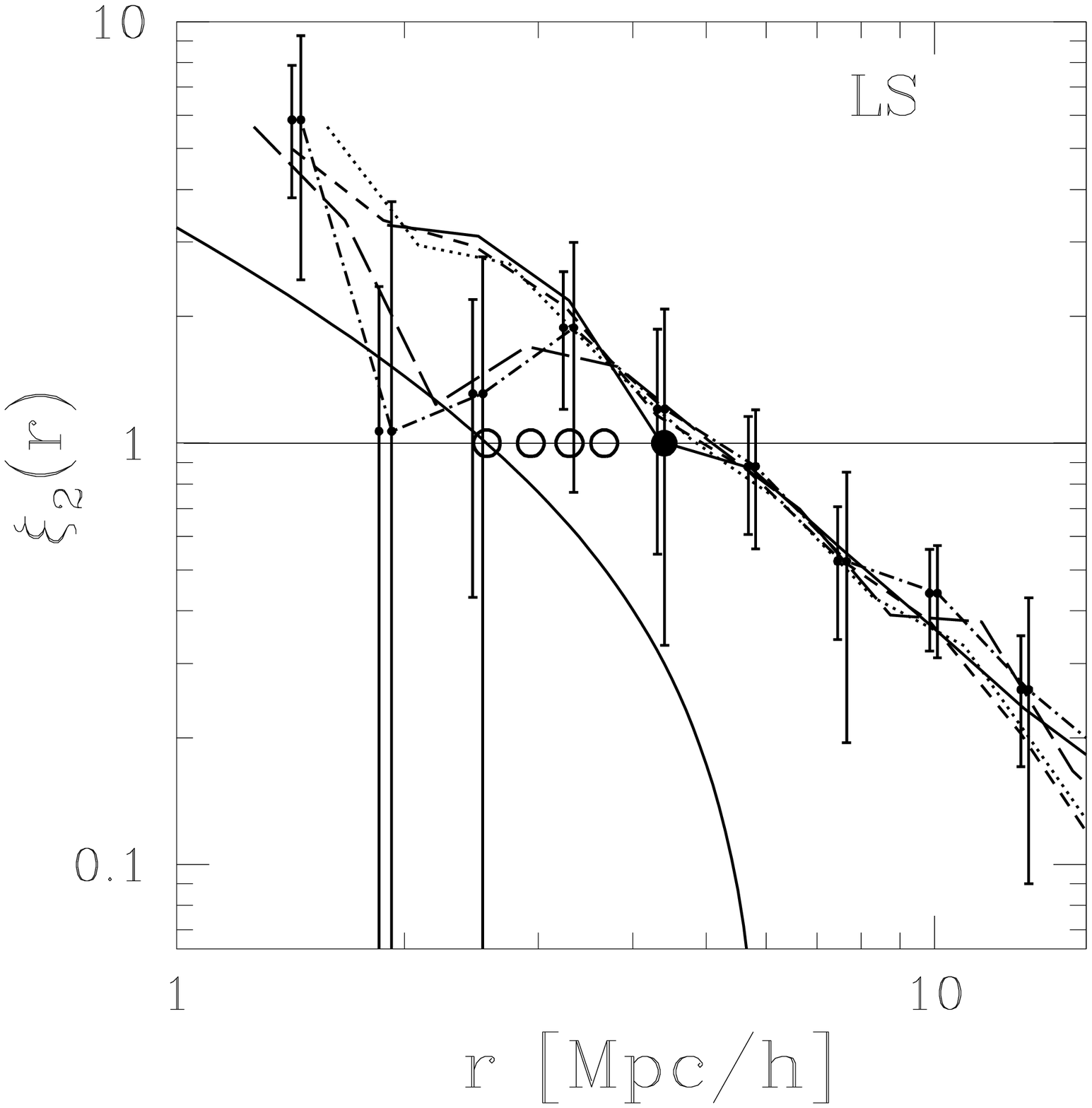}
\caption{$\xi(r)$ for a sequence of volume-limited samples with
changing limiting luminosity $L_{\rm lim}$ from the PSCz determined
with the Landy \& Szalay estimator with marks similar to
Fig.~\ref{fig:ssrs2-vollim}. The smooth solid line is $\xi(r)$ for a fractal  with $D=2.2$ 
and $r_0(70\hMpc)$ according to Eq.\eqref{eq:fractal_r0}.
Two-$\sigma$ error bars are shown 
only for the 120\hMpc\ sample. 
The larger error bars were determined from the mock catalogues. 
The Poisson error bars are slightly shifted to the left.}
\label{fig:pscz-vollim}
\caption{$\xi(r)$ for samples with varying depth but with the same
limiting Luminosity $L_{\rm lim}$ extracted from a volume-limited
sample of the PSCz with a depth of $R_{\text{max}}=120\hMpc$ similar to Fig.~\ref{fig:ssrs2-seq}. 
Marks as in Fig.~\ref{fig:pscz-vollim}, but the two-$\sigma$ error bars are shown 
only for the 70\hMpc\ sample.
}
\label{fig:pscz-seq}
\end{figure}

Both the galaxies in the SSRS2 and the CfA2 were selected in the
optical waveband. The galaxies in the PSCz survey were selected according to
their flux in the infrared as detected by the IRAS satellite.  A
detailed description of the IRAS PSCz galaxy catalogue may be found in
{\cite{saunders:pscz}}.  I extract volume-limited samples from the
PSCz survey using luminosity distances within the standard masked
area.
I approximate the
sample geometry by two spherical caps with galactic latitude
$b\ge5^\circ$ for the northern part and with $b\le-5^\circ$ for the
southern part.  Hence, I neglect some regions which were left empty
due to galactic absorption or confusion in the IRAS PSC maps.  I
filled these empty regions with random points assuming the same number
density as in the fully sampled region. No differences in the
correlation properties between the filled and unfilled samples are
visible in the two-point measures.
The $\xi(r)$ determined from a sequence of volume-limited samples is 
inconsistent with the fractal prediction and  
shows no significant variation of $r_0$ with the sample size
(Fig.~\ref{fig:pscz-vollim}).  As expected, extracting subsamples from
one volume-limited sample with $R_{\text{max}}=120\hMpc$ does not
change this behaviour, although the fluctuations increase in the 
sparser samples (Fig.~\ref{fig:pscz-seq}).  Clearly, there
is neither an indication for a fractal scaling of $r_0$ with the
sample depth nor for luminosity segregation (see also
{}\cite{szapudi:correlationspscz}).
Due to the selection of galaxies in the infrared one does miss early
type (e.g.\ elliptical) galaxies. Because of this selection one  does not
find luminosity segregation in the PSCz {}\cite{beisbart:luminosity}.
To go beyond the error estimates relying on the Poisson process I estimate the 
errors for $\xi(r)$ from the fluctuations 
between eleven mock galaxy catalogues, constructed from an $N$-body simulation based on a $\Lambda$CDM cosmology\footnote{
A description of the procedure and references to articles describing the simulation and the construction of the mock catalogues can be found in {}\cite{kerscher:pscz}. }.
For such a clustered point distribution, the error bars are 
larger than in the case of the Poisson process, but still of the same order (see 
Figs.~\ref{fig:pscz-vollim},\ref{fig:pscz-seq}). In both models, 
these errors are smaller than the predicted effects of fractal scaling on $\xi(r)$ (two-$\sigma$ 
error bars are shown in the plots).

\section{Summary}

By analysing three different galaxy catalogues I could show that the
amplitude of the correlation function $\xi(r)$ is depending on the
luminosity. Luminosity segregation already has been found in the
SSRS2, CfA2 and also the 2dF and the SDSS galaxy catalogues (see e.g.\
{}\cite{hamilton:evidence,jones:multifractal,martinez:galaxy,benoist:biasing,willmer:southern,norberg:2df,zehavi:galaxy}). In
these investigations mainly volume-limited samples with varying
depths or directly flux-limited samples have been used.  Using such 
samples one is not able to separate the
influence of luminosity segregation from a possible fractal scaling.
In samples with a fixed lower limit in the luminosity 
and then reducing the depth of the samples I found no
significant changes in the correlation function.  
Specifically, I found no sign for a growth of the correlation length $r_0$ with 
increasing sample depths in these cases.
This is a strong indication that the growth of $r_0$ in standard
volume-limited samples is caused by luminosity segregation, and a
fractal explanation is disfavoured.
Using a similar construction {}\cite{labini:scale} found a growing
$r_0$ in the Perseus Pisces survey (PPS).  Our consistent results,
both from the significantly larger SSRS2 and the CfA2 indicate that
the PPS is too small in size to give reliable results.  In the CfA2
{}\cite{martinez:does} found also no significant growth of $r_0$ using
a comparable method.
{}\cite{beisbart:luminosity} used mark correlation functions to quantify 
the luminosity dependency of the galaxy clustering in the SSRS2 in a scale
dependent way, uninfluenced by inhomogeneities.

I limited my investigations to the question: what causes the growing
amplitude of the two-point correlation function?  Already with the
current galaxy catalogues one is able to show that the amplitude of the
two-point correlation function is depending on the luminosity of the
galaxies. 
If one a priori assumes that the galaxy distribution
is a fractal, these results may be interpreted as large fluctuations,
which indeed are common in fractals. However, these fluctuations must
conspire in all the samples considered here, to give the
observed result of a constant amplitude of the two-point correlation function.
Fluctuations in the morphological properties of the large scale distribution of 
galaxies have been detected out to a scale of  200\hMpc\  {}\cite{kerscher:pscz}. 
However, these fluctuations are barely visible with two-point 
measures, and they are still compatible with fluctuations expected from 
a $\Lambda$CDM model.
I did not comment on the topic, whether one already does see a
turnover to homogeneity from current galaxy surveys
{}\cite{kerscher:pscz}, and whether $\xi(r)$ or $\Gamma(r)$ shows the
more extended scaling regime. For a discussion see
{}\cite{mccauley:thegalaxy,gaite:fractal}.  I expect that the
completed Sloan Digital Sky Survey will offer conclusive evidence for
these points.

\acknowledgments
I would like to thank Ruth Durrer and Francesco Sylos Labini for
organising the workshop "Facts and Fiction in Cosmology" in April 2001
in Sils Maria. The discussions during this workshop generated the
impulse to re-investigate these galaxy catalogues.
For comments on an earlier version of the manuscript I am grateful to Alvaro Dom{\'\i}nguez,
Jose Gaite, Michael Joyce, Marco Montuori and especially to Claus
Beisbart.
I acknowledge support from the {\em Sonderforschungsbereich 375
f{\"u}r Astroteilchenphysik der DFG}.


\end{document}